\documentclass[11pt]{article}
\pdfoutput=1
\usepackage[utf8]{inputenc}
\usepackage{dsfont}
\usepackage{diagbox}
\usepackage{amsfonts}
\usepackage{mathtools}
\allowdisplaybreaks[4]        
\usepackage{amssymb}
\usepackage{euscript}     
\usepackage{braket}
\usepackage{starfont}
\usepackage{color,soul}         
\usepackage{braket}
\usepackage{tensor}        
\usepackage{amsthm}
\usepackage{graphicx}
\usepackage{slashed}
\usepackage{leftidx}
\usepackage{subfigure}
\usepackage{bbm}
\usepackage{indentfirst}

\definecolor{outerspace}{rgb}{0.25, 0.29, 0.3}
\definecolor{scarlet}{rgb}{1.0, 0.13, 0.0}
\usepackage[header,title,page,titletoc]{appendix}  
\definecolor{princetonorange}{rgb}{1.0, 0.56, 0.0}
\definecolor{WildStrawberry}{rgb}{1.0, 0.26, 0.64}
\definecolor{rossocorsa}{rgb}{0.83, 0.0, 0.0}
\definecolor{navyblue}{rgb}{0.0, 0.0, 0.5}
\usepackage[numbers,sort&compress]{natbib}  
\usepackage{float}
\usepackage[paper=a4paper,margin=1in]{geometry}
\usepackage{titlesec}

\pdfoutput=1
\usepackage{amsmath}
\usepackage{color}
\usepackage{amsfonts}
\usepackage{amssymb}
\usepackage{graphicx}
\usepackage{geometry}
\usepackage{amssymb,epsfig,subfigure}
\usepackage{amssymb}
\usepackage{hyperref}
\usepackage{comment}
\usepackage[font=footnotesize]{caption}
\usepackage[T1]{fontenc}
\usepackage[utf8]{inputenc}
\usepackage{latexsym}

\usepackage{cancel}

\usepackage[numbers,sort&compress]{natbib}  
\usepackage{float}

\usepackage{dsfont}
\usepackage{diagbox}

\usepackage{mathtools}
\allowdisplaybreaks[4]

\usepackage{euscript}     
\usepackage{braket}
\usepackage{starfont}
\usepackage{color,soul}         
\usepackage{tensor}        
\usepackage{amsthm}

\usepackage{slashed}
\usepackage{leftidx}
\usepackage{bbm}

\makeatletter
\renewcommand\section{\@startsection {section}{1}{\z@}%
                                 {-3.5ex \@plus -1ex \@minus -.2ex}
                                   {2.3ex \@plus.2ex}%
                                   {\normalfont\large\bfseries}}
\renewcommand\subsection{\@startsection{subsection}{2}{\z@}%
                                   {-3.25ex\@plus -1ex \@minus -.2ex}%
                                     {1.5ex \@plus .2ex}%
                                     {\normalfont\bfseries}}
\renewcommand\subsubsection{\@startsection{subsubsection}{3}{\z@}%
                                   {-3.25ex\@plus -1ex \@minus -.2ex}%
                                     {1.5ex \@plus .2ex}%
                                     {\normalfont\itshape}}
\makeatother

\def\pplogo{\vbox{\kern-\headheight\kern -29pt
\halign{##&##\hfil\cr&{\ppnumber}\cr\rule{0pt}{2.5ex}&\ppdate\cr}}}
\makeatletter
\def\ps@firstpage{\ps@empty \def\@oddhead{\hss\pplogo}%
  \let\@evenhead\@oddhead 
}
\def\maketitle{\par
 \begingroup
 \def\thefootnote{\fnsymbol{footnote}}
 \def\@makefnmark{\hbox{$^{\@thefnmark}$\hss}}
 \if@twocolumn
 \twocolumn[\@maketitle]
 \else \newpage
 \global\@topnum\z@ \@maketitle \fi\thispagestyle{firstpage}\@thanks
 \endgroup
 \setcounter{footnote}{0}
 \let\maketitle\relax
 \let\@maketitle\relax
 \gdef\@thanks{}\gdef\@author{}\gdef\@title{}\let\thanks\relax}
\makeatother

\numberwithin{equation}{section}

\newcommand\eea{\end{eqnarray}}
\newcommand\bea{\begin{eqnarray}}

\def\beq{\begin{equation}}
\def\eeq{\end{equation}}

\newcommand{\be}{\begin{equation}}
\newcommand{\ee}{\end{equation}}
\newcommand{\ba}{\begin{align}}
\newcommand{\ea}{\end{align}}
\newcommand{\bg}{\begin{gather}}
\newcommand{\eg}{\end{gather}}
\newcommand{\bseq}{\begin{subequations}}
\newcommand{\eseq}{\end{subequations}}

\usepackage{hyperref}
\hypersetup{
    colorlinks,
    citecolor=rossocorsa,
    filecolor=navyblue,
    linkcolor=navyblue,
    urlcolor=navyblue
}

\begin{document} 

\begin{titlepage}

\begin{center}

\phantom{ }
\vspace{3cm}

{\bf On completeness and generalized symmetries in quantum field theory}
\vskip 0.5cm
Horacio Casini${}^{*}$,  Javier M. Mag\'an${}^{\ddagger}$
\vskip 0.05in
\small{${}^{*}$ \textit{Instituto Balseiro, Centro At\'omico Bariloche}}
\vskip -.4cm
\small{\textit{ 8400-S.C. de Bariloche, R\'io Negro, Argentina}}
\vskip -.10cm
\small{${}^{\ddagger}$ \textit{David Rittenhouse Laboratory, University of Pennsylvania}}
\vskip -.4cm
\small{\textit{ 209 S.33rd Street, Philadelphia, PA 19104, USA}}
\vskip -.10cm
\small{${}^{\ddagger}$ \textit{Theoretische Natuurkunde, Vrije Universiteit Brussel (VUB) }}
\vskip -.4cm
\small{\textit{ and The International Solvay Institutes}}
\vskip -.4cm
\small{\textit{ Pleinlaan 2, 1050 Brussels, Belgium}}

\begin{abstract}

We review a notion of completeness in QFT arising from the analysis of basic properties of the set of operator algebras attached to regions. In words, this completeness asserts that the physical observable algebras produced by local degrees of freedom are the maximal ones compatible with causality. We elaborate on equivalent statements to this completeness principle such as the non-existence of generalized symmetries and the uniqueness of the net of algebras. We clarify that for non-complete theories, the existence of generalized symmetries is unavoidable, and further, that they always come in dual pairs with precisely the same ``size''. Moreover, the dual symmetries are always broken together, be it explicitly or effectively.   Finally, we comment on several issues raised in recent literature, such as the relationship between completeness and modular invariance,  dense sets of charges, and absence of generalized symmetries in the bulk of holographic theories.

\end{abstract}
\end{center}

\small{\vspace{5 cm}\noindent${}^{\text{\text{*}}}$casini@cab.cnea.gov.ar\\
${}^{\dagger}$magan@sas.upenn.edu\\}

\end{titlepage}

\setcounter{tocdepth}{2}

\newpage

\section{Introduction}

In an article presented at the 2002 Centennial Symposium, Polchinski formulated a pair of ``completeness'' principles, namely \cite{Polchinski:2003bq}:

\bigskip

\noindent {\sl in any theoretical framework that requires charge to be quantized, there will exist magnetic monopoles,}

\bigskip

\noindent and, more strongly,
 
\bigskip

\noindent {\sl in any fully unified theory, for every gauge field the lattice of electric and magnetic charges is maximal.}

\bigskip

This idea of completeness expresses the necessity of counting with a large enough set of operators. More recently, the idea has been taken up and developed in different directions in the context of effective theories of quantum gravity (see for example \cite{Banks:2010zn}).
The traditional arguments supporting these statements come from the physics of black holes. They are related to the possibility of black hole evaporation, problems with remnants, and the existence of black hole solutions with non-trivial gauge charges. Hence, the idea is that a certain ``completeness''  is enforced on a quantum field theory (QFT) that is consistently coupled to gravity, though this completeness is not a logical requirement for  QFT in itself.  

Stepping back, as a preliminary step in this research, it would be desirable to understand in more clear terms what is the scope of the concept of completeness in a QFT without gravity. In the literature, starting with \cite{Banks:2010zn}, this issue has rightly been connected with symmetries and the existence of a sufficient number of charged fields. However, many different models, phenomena, and symmetries appear in different dimensions, and there is a lack of a unifying principle as well as a clear connection between the ideas of completeness on one hand, and that of symmetries on the other.  Relatedly, we have not found in the literature a notion of completeness that is a priori independent of the existence of symmetries. The main objective of this article is to review a formulation of a precise proposal for completeness in the most general context and to understand the connections with symmetries that arise from relaxing this requirement.

With this objective in mind, it is of obvious interest to formulate those principles in QFT with as few assumptions as possible. For example, we do not want to assume small coupling or the existence of particular types of operators. To this end, we step on the approach to ``generalized symmetries'' and ``completeness'' in QFT developed recently in Ref.\cite{Casini:2020rgj}, which we review and deepen. This non-lagrangian/model-independent approach is based on probably the most basics properties of any QFT,\footnote{By ``QFT'' we also include local effective field theories valid until certain cutoff or QFT's which are defined through lattice completions. We hope it will become clear that the approach and properties described here are in a precise sense ``macroscopic''. They do not depend on the UV completion/regularization of the model.} which are those of the existence of local degrees of freedom and causality. The analysis of these properties leads to a unification of various ideas about completeness into a simple one, which is codified in the following statement, to be formulated mathematically below:

\newpage

\noindent {\sl A QFT is complete if for an arbitrary region of the space the observable algebra generated by the local degrees of freedom is the maximal one compatible with causality.}

\bigskip

Besides, as proposed and developed in \cite{Casini:2020rgj}, a violation of the previous property entails the existence of a generalized symmetry. Such generalized symmetries are characterized by the topology of the regions $R$ in which the previous assertion is violated. Generalized global symmetries, as defined in \cite{Gaiotto:2014kfa}, are included as special cases.

While the present approach shows the precise equivalence between non-completeness and the existence of symmetries,\footnote{As mentioned before, the intuition behind this connection has appeared in previous literature on the subject. One can find it already in \cite{Banks:2010zn}, and more recently in \cite{Rudelius:2020orz,Casini:2020rgj,Heidenreich:2021tna}. But several problems have been pointed out, in particular in regards to discrete gauge theories, see \cite{Harlow:2018tng,Rudelius:2020orz,Heidenreich:2021tna} for discussions on this point. We will see below how the present approach clarifies these issues.} it also has some quite direct, simple, but important new consequences. In this context, it is straightforward to show that generalized symmetries always come in ``dual pairs'', as dictated by Von Neumann's double commutant theorem. Moreover, in a way that can be made very precise, the amount of ``electric symmetry'' is bound to be the same as the amount of ``magnetic symmetry''. In particular, this includes the idea that explicit or effective disappearance of an electric symmetry implies the same for the magnetic counterpart, and we describe how this happens. These consequences are expected to lead to new insights in the context of quantum gravity, as we remark in the last section.
 
In the way, we will also describe the differences between two notions of completeness which might cause some confusion. One is related to the existence of maximal nets of non-local operators in pure gauge theories, dubbed Haag-Dirac nets in \cite{Casini:2020rgj}, and the other is the more interesting completeness alluded above and conjectured to hold for theories of quantum gravity. We will also clarify some problems that appear when considering dense sets of charges, and and comment on the absence of generalized symmetries in the bulk of holographic theories.

\section{Basic structures of QFT and completeness}
\label{three}
In a QFT the spacetime is inscribed into the quantum theory through the existence of algebras of operators ${\cal A}(R)$ attached to spacetime regions $R$.  The operators in the algebras ${\cal A}(R)$ can be thought of as idealized laboratory operations that can be performed in $R$. As such we can think of these algebras as formed by combinations of observables in $R$. This description applies to lattice theories as well, but in a continuum theory, it is mathematically convenient that the algebras are endowed with some topological property. A physically motivated topology is given by the idea that having a sequence of operators we could as well think we have another operator which cannot be distinguished from elements of this sequence by any finite but arbitrary number of experiments, with any fixed, but arbitrary, precision. This is called the weak topology for operators, and the algebras closed under this topology are von Neumann algebras. Hence, this is a mathematically convenient but physically irrelevant choice, similar in spirit to the choice of real numbers instead of rational ones in the description of the space. von Neumann algebras (that include all finite-dimensional algebras of operators in Hilbert space) are nicely characterized by an algebraic property. If we call ${\cal A}'$ to the set of operators that commute with ${\cal A}$, a von Neumann algebra is any set of operators satisfying 
\be   
{\cal A}''={\cal A}. 
\ee
This is von Neumann's double commutant theorem.

In a QFT a natural expectation is that the operator content of the theory is ultimately generated by local degrees of freedom, or field operators, localized in arbitrarily small regions of space. To give a mathematical form for this ``additivity'' principle, without assuming any particular QFT model, we can construct algebras for any region $R$ by generating them with other smaller algebras associated with balls contained in $R$. As a primitive form of additivity, we will assume that the algebra of a ball coincides with the algebra generated by any collections of smaller balls covering it.\footnote{A conformal generalized free field with a two-point function $|x-y|^{-2 \Delta}$ will not satisfy this primitive form of additivity if we associate to a spatial ball the algebras of its associated causal double cone in spacetime, see \cite{Duetsch:2002hc}. This type of field appears at large $N$ limits but disappears when finite $N$ corrections are considered, see \cite{ElShowk:2011ag} for an extensive discussion. The problem with these theories is better described by the absence of local causal evolution, or the absence of stress tensor, rather than by a lack of additivity.} To be more precise, this suggests we can construct an {\sl additive net} by assigning to any region $R$ (with any given topology) the following {\sl additive} algebra
\be
{\cal A}_{\textrm{add}}(R)\equiv \bigvee_{B \, \textrm{ball}\,, \cup B=R} {\cal A}(B)\,. 
\ee  
Here the symbol ${\cal A}_1\vee {\cal A}_2=({\cal A}_1\cup {\cal A}_2)''$ means the algebra generated by the two.  More generally, a net is an assignation of algebras to any region of the space. The additive net just defined is an example that is always present and has, by definition,  the additive property
\be
{\cal A}_{\rm add}(R_1\cup R_2)={\cal A}_{\rm add}(R_1)\vee {\cal A}_{\rm add}(R_2)\,.\label{3}
\ee 

 In intuitive terms, if we were to construct the best laboratory supported on a given region $R$, the additive algebra would be the set of measurements available to such laboratory. For example, as we describe in more detail later, in absence of electric charges, a ring-like laboratory would not be able to measure certain Wilson loops going along the ring. It would be able to measure such  Wilson loops whenever they do not wrap the ring since in that case, they are magnetic fluxes belonging to the additive algebra of the region where the laboratory is supported. In a precise sense, the additive algebra ${\cal A}_{\rm add}(R)$ is the minimal one that can be associated with a region in a given QFT, and the additive net is the minimal net. Notice that this axiom implies another one typically seen in these discussions, which is that of isotonia
\be
{\cal A}_{\rm add}(R_1)\subseteq  {\cal A}_{\rm add}(R_2)\,,\hspace{1cm}R_1\subseteq R_2\,. \label{isotonia}
\ee
Without this ordering relation, the idea that an operator belongs to a certain region loses its meaning. 

A second fundamental property in relativistic QFT is {\sl causality}
\be
{\cal A}_{\textrm{add}}(R) \subseteq  ({\cal A}_{\textrm{add}}(R'))'\,,\label{causality}
\ee
where $R'$ is the causal complement of $R$, i.e. the set of points spatially separated from $R$. We recall that  ${\cal A}'$ is the algebra of all operators that commute with those of ${\cal A}$. This property just embodies the commutativity of operators at spacial distance. 

Looking at the causality principle, we may naturally wonder whether the inclusion of algebras (\ref{causality}) is saturated or not. If it is not saturated for a certain region $R$, we must conclude there are at least two natural algebras associated with the same region $R$, namely ${\cal A}_{\textrm{add}}(R)$ and $({\cal A}_{\textrm{add}}(R'))'$. Given the previous causality requirement, it is clear that $({\cal A}_{\textrm{add}}(R'))'$ is the largest possible algebra associated with $R$ which still commutes with the observables in $R'$.  We will thus rename it as
\be
{\cal A}_{\textrm{max}}(R)\equiv({\cal A}_{\textrm{add}}(R'))' \;. 
\ee

The failure of saturation in (\ref{causality}) entails a certain lack of operators in the locally generated algebras (the additive algebras) such that some other operators could be added without violating the causality principle. 
This observation was used in \cite{Casini:2020rgj} to suggest a  definition of completeness in QFT, namely
\be\label{Hcom}
{\cal A}_{\textrm{add}}(R) = ({\cal A}_{\textrm{add}}(R'))'= {\cal A}_{\textrm{max}}(R)\equiv {\cal A}(R)\,,\hspace{.4cm} \forall R \,, \hspace{.7cm}\textrm{(complete theory)} \,.
\ee 
As we will elaborate below, this idea of ``completeness'', simple and general as it is, makes precise the intuition about completeness that has been previously elaborated in the literature.  
 
 Notice that in complete theories there is no ambiguity for the algebra associated with a given region. This algebra can only be the minimal one, namely the additive algebra, which coincides with the maximal one. The only possible net of algebras is the additive net.  
For a ball, the property ${\cal A}(B)={\cal A}(B')'$ is called Haag duality. This property is expected to be true for QFT's satisfying certain minimal requirements. For a more general region with non-trivial topology, the property ${\cal A}(R)={\cal A}(R')'$ is usually called duality. The present completeness criterium then asks for duality of the additive observable algebras. This leads to a more precise form of our verbal definition in the introduction, since we can express the idea of completeness in two additional ways:
 \bigskip

\noindent \hspace{1.3 cm}{\sl Completeness $\equiv$ uniqueness of the net $\equiv$ duality for the additive observable net.}

\bigskip

We remark this definition makes perfect sense for non-relativistic or lattice theories because we have only talked about algebras associated with regions of the space. For relativistic theories, the completeness postulate becomes even more elegant mathematically. This was called ``tentative postulate'' in Haag's book \cite{haag2012local}. Both for algebras and spacetime regions there exist the relations and logical operations $(\subseteq, \wedge, \vee, ')$, where $\wedge$ is the intersection of algebras or regions respectively, and $\vee$ between two regions is the causal completion of the union , $R_1\vee R_2=(R_1\cup R_2)''$. Then, the completeness criterium in the relativistic setting demands that the two structures are mapped to each other by the map $R\rightarrow {\cal A}(R)$ that defines the QFT, as in eqs. (\ref{3}) and (\ref{Hcom}). Mathematically this is a homomorphism of orthocomplemented lattices, see \cite{haag2012local}. From the mathematical point of view, it is plain that the completeness postulate is about having the most harmonious possible relation between algebras regions.

\section{Incompleteness as the origin of dual generalized symmetries.  }
\label{ggs}
  The violation of the duality property of the additive algebra, i.e. the non-completeness of the net of algebras of the QFT, is directly related to the appearance of generalized symmetries  \cite{Casini:2020rgj}. 
To see this, consider a region such that
\be
{\cal A}_{\textrm{add}}(R) \subsetneq  {\cal A}_{\textrm{max}}(R) \;.\label{rala}
\ee 
We have then two canonical algebras associated with such a region. Since one is included in the other, there must be a set $\{a\}$ of ``non-locally generated operators'' in the region $R$ such that
\be 
{\cal A}_{\textrm{max}}(R)={\cal A}_{\textrm{add}}(R)\vee \{a\}\;.
\ee
 This does not imply a tensor product structure. In fact the non-local operators ${a}$ do not commute with ${\cal A}_{\textrm{add}}(R)$. To ease the language, we will simply call the operators $a$ the ``non-local operators'' in $R$, because they cannot be generated in a local manner inside $R$. In the literature, these types of operators are sometimes called topological operators. In what sense they can be considered topological will be deal with shortly. For the moment we keep the name ``non-local operators'', to remind us that their origin is the violation of duality in the associated region.
 
It is important to notice that being a non-local operator is relative to a certain region since although one operator $a$ can be non-local in a certain region $R$, the same operator might become locally generated in a bigger region $\tilde{R}$,  $R \subset \tilde{R}$. In particular, if the algebra ${\cal A}_{\rm add}(\tilde{R})$ satisfies duality, all operators which are confined to $\tilde{R}$ by causality are locally generated in $\tilde{R}$. If we have Haag duality for the additive net any operator is locally generated inside balls. 

The operators $a$ may now be used to define irreducible classes/sectors $[a]$ of operators in ${\cal A}_{\textrm{max}}(R)$. The class $[a]$ of ${\cal A}_{\textrm{max}}(R)$ is the set of operators of the form $\sum_\lambda O_1^\lambda \,a\, O_2^\lambda$, with $O_1^\lambda$ and $O_2^\lambda$ being locally generated operators in $R$, i.e operators belonging to ${\cal A}_{\textrm{add}}(R)$. A sector $[a]$ is said to be irreducible if there are no non-trivial subspaces inside $[a]$ that are invariant under the simultaneous left and right action of the additive algebra. In the opposite case it is said to be reducible and it can be decomposed into a sum of irreducible ones. The class $[1]$ coincides by definition with ${\cal A}_{\textrm{add}}(R)$. Its fusion rules are by construction $[1][a]=[a][1]=[a]$. Because ${\cal A}_{\textrm{max}}(R)$ is an algebra, the set of classes must close a fusion algebra between themselves $[a][a']=\sum_{a''} [n]^{a''}_{a a'} [a'']$. These fusion rules simply indicate which classes appear in the decomposition. These fusion rules might be related to the fusion of representations or conjugacy classes of a certain group, but they can be more general, the paradigmatic example being QFT's in two dimensions. We comment more on the potential range of possibilities below.

Now, the non trivial inclusion (\ref{rala}), forces the algebra of the complement $R'$ of $R$ to violate duality as well. The reason is simple, it just follows from Von Neumann's double commutant theorem  ${\cal A}''={\cal A}$. We can canonically associate two algebras to the complementary region, namely 
\bea 
{\cal A}_{\textrm{add}}(R')&=&{\cal A}_{\textrm{max}}(R)'\,,\nonumber \\
{\cal A}_{\textrm{max}}(R')&=&{\cal A}_{\textrm{add}}(R)'\;.
\eea
The second of these algebras clearly contains the first. But, given (\ref{rala}), this inclusion has to be strict, and the two algebras cannot coincide. Indeed, the equality of  ${\cal A}_{\textrm{add}}(R')$ and  ${\cal A}_{\textrm{max}}(R')$ entails the one of the algebras corresponding to $R$
\bea 
&&{\cal A}_{\textrm{add}}(R')={\cal A}_{\textrm{max}}(R') \iff {\cal A}_{\textrm{max}}(R)'={\cal A}_{\textrm{add}}(R)' 
\\
&&\hspace{2cm}\iff ({\cal A}_{\textrm{max}}(R)')'=({\cal A}_{\textrm{add}}(R)')'\iff {\cal A}_{\textrm{max}}(R)={\cal A}_{\textrm{add}}(R)\;.\nonumber
\eea
We conclude that if there are non local operators $\{a\}$ for $R$ there must be non local operators $\{b\}$ for the complement $R'$, 
 \be 
{\cal A}_{\textrm{max}}(R')={\cal A}_{\textrm{add}}(R')\vee \{b\}\;.
\ee
These non local operators must have their own fusion rules $[b][b']=\sum_{b''} [n]^{b''}_{b b'} [b'']$. 

Because  ${\cal A}_{\textrm{max}}(R)={\cal A}_{\textrm{add}}(R')'$ the non local operators of $R$ commute with the additive operators in $R'$ and viceversa. However, the non local operators $\{a\}$ and $\{b\}$ for $R$ and $R'$ cannot commute (all of them) with each other. Such commutation would imply ${\cal A}_{\textrm{max}}(R)\subseteq ({\cal A}_{\textrm{max}}(R'))'={\cal A}_{\textrm{add}}(R)$, which is not possible if the inclusions are strict.   
The failure of commutativity between operators located at a space-like distance is typically a sign of a violation of causality. We want to remark that this is not the case here. Although the non-local operators $a$'s and $b$'s are naturally attached to complementary regions $R$ and $R'$ (because they commute with the additive algebras in their respective complementary regions), they cannot be constructed locally there. In most cases, we can construct them locally in bigger regions, and these will have a non-trivial intersection with each other, explaining the failure of commutativity.

The algebras of non-local operators for each region can be seen as generators of transformations of the maximal algebras of the complementary regions, see \cite{Casini:2020rgj} and below for specific instances. In this sense, the non-local operators act as generalized symmetry operations, where the charged objects are the complementary non-local operators and the neutral objects are the observables in the additive algebra. Examples include the generalized global symmetries defined in \cite{Gaiotto:2014kfa}. But we expect this more abstract description will lead to discover and unify more exotic phenomena, including exotic fusion rules and more complicated violations of duality.

To summarize, an important conclusion is drawn from these simple observations. It can be stated as a new characterization of completeness 
\bigskip

\noindent \hspace{3.5 cm} {\sl completeness $\equiv$ absence of generalized symmetries.}

\bigskip

In \cite{Magan:2020ake,Casini:2020rgj}, the previous duality structure was summarized by a quantum complementarity diagram
\bea\label{cdiaor}
{\cal A}_{\textrm{add}}(R)\vee \{a\} & \overset{E}{\longrightarrow} &{\cal A}_{\textrm{add}}(R)\nonumber \\
\updownarrow\prime \!\! &  & \,\updownarrow\prime\\
{\cal A}_{\textrm{add}}(R')& \overset{E'}{\longleftarrow} & {\cal A}_{\textrm{add}}(R')\vee \{b\}\,.\nonumber 
\eea
In this diagram, on the upper side, we have the two algebras associated with region $R$. They are connected by a linear and positive projection map $E$, which in the algebraic context is called a conditional expectation. In this case, this projection kills the non-local operators ${a}$. For a continuum theory, this map is expected to be unique. Going down in the diagram amounts to taking commutants, producing the algebras naturally associated with the complementary region $R'$. These are connected by a dual conditional expectation $E'$ that kills the non-local operators $b$.

When one of the generalized symmetries, generated by an algebra of non-local operators, represents a symmetry group, this conditional expectation is just defined as an average over the group. This kills any dual charged operators. But this ``average over the group'' can be naturally generalized by the present conditional expectations. If a conditional expectation kills the non-local operators in a region, then it is generated by the algebra of dual non-local operators, see \cite{Longo:1994xe} for a thorough exposition. This ``generalized symmetry action'', as described before, tells us that the additive algebras are the neutral objects of these ``generalized symmetries''.

Now we observe that having the upper part in the complementarity diagram~(\ref{cdiaor}) implies having the lower part. There is no other way around. This leads to our second important conclusion of this section:
 \bigskip

\noindent \hspace{3.5cm} {\sl generalized symmetries always come in dual pairs.}

\bigskip

Not only generalized symmetries come in dual pairs, but also the ``amount'' or ``size'' of these dual symmetries is, in a precise sense, the same. The concept of ``size'' might not appear clear, since the algebra ${\cal A}_{\textrm{max}}(R)={\cal A}_{\textrm{add}}(R)\vee {a}$ is not a tensor product, and the algebras are of infinite type. These types of questions motivated Jones to introduce the so-called ``Jones index'' in his seminal work \cite{Jones1983}. In such work, the index was defined for type II algebras, but shortly after Kosaki \cite{KOSAKI1986123} and Longo \cite{longo1989} independently generalized this notion to all types of algebras, including the ones appearing in QFT. The Jones index is a precise measure of the size of the algebra of non-local operators, or, equivalently, the size of the inclusion ${\cal A}_{\textrm{add}}(R)\subset {\cal A}_{\textrm{max}}(R)$.  See \cite{teruya,giorlongo,Magan:2020ake} for simpler introductions to the concept and specific computations. When applied to symmetries related to groups, the index just becomes the number of elements in the group. When the index is finite, the dual inclusions ${\cal A}_{\textrm{add}}(R)\subset {\cal A}_{\textrm{max}}(R)$ and ${\cal A}_{\textrm{add}}(R')\subset {\cal A}_{\textrm{max}}(R')$ have the same index. For complete theories, the index is one. See \cite{Longo:1994xe} for a review and references.\footnote{It is interesting to observe that this duality, which includes electromagnetic duality, is seen in this light as akin to the fundamental duality in logic.  This connection arises from the fact that the orthocomplemented lattice structure of both, subalgebras and causal regions, is shared by the structure of propositions in both Boolean classical logic and quantum logic \cite{casini:hal-00129295}.}

It is also worth noting that expectation values of non-local operators are natural order parameters for the corresponding generalized symmetries. In particular, they are the only operators which can display an area law, and hence, for example, act as an order parameter for confinement in non-Abelian gauge theories \cite{Casini:2020rgj}. 

\section{Some examples}
The violation of duality directly leads to the existence of classes/sectors of operator algebras $[a]$ and $[b]$ with non-trivial fusion rules.  Very little is known about the allowed fusion algebras for different $R,  R'$ and spacetime dimensions. One natural assumption, which is the case in most of the examples, is that the sectors are {\sl transportable}. This means that given a continuous deformation from $R_1$ to $R_2$ the corresponding sectors $a_1$ and $a_2$ can be identified as belonging to the same sector $a$ of the tube of homotopy between the two deformed regions. The ``transport'' from $a_1$ to $a_2$ is made by the action of locally generated operators in this tube of homotopy. In this case, the algebra of non-local operators $\left\lbrace a\right\rbrace $ and the associated generalized symmetry can be considered ``topological''. More concretely, the classes, fusion rules, and commutation relations for the dual non-local operators depend only on the topology of $R$. However, the possibility of non-transportable sectors cannot be discarded. It seems to be realized by the so-called fractons, in theories without relativistic symmetry, see the review \cite{Pretko_2020} and the recent proposals for continuum QFT \cite{Seiberg:2020bhn,Seiberg:2020wsg,Seiberg:2020cxy}. We will assume transportability in what follows.

Let us review now some examples. Let us start with global symmetries, and to simplify assume there is no spontaneous symmetry breaking. In this scenario, the natural additive algebra of observables is generated by the neutral operators ${\cal A}(B)$ in balls. In particular, there is no access to charged operators in a laboratory. Consider a region $R$ formed by union of two balls $R=B_1\cup B_2$. Let $\psi^{i,r}_1$, $\psi^{i,r}_2$ be charge creating operators in $B_1$ and $B_2$, corresponding to the irreducible representation $r$, and where $i$ is an index of the representation. From the present perspective, these operators do not belong to the observable algebra (the neutral QFT or orbifold theory). But using such operators we can construct a neutral operator, the {\sl intertwiner} corresponding to this representation
\be\label{Ir}
{\cal I}_r= \sum_i \psi^{i,r}_1 (\psi^{i,r}_2)^\dagger\;.
\ee  
This is invariant under global group transformations and it is, therefore, an observable. It commutes with operators in the complementary region $R'=(B_1\cup B_2)'$, but it cannot be generated additively by operators in the neutral algebras ${\cal A}(B_1)$ and ${\cal A}(B_2)$. 
Nonetheless, it is an operator of the QFT which can be additively generated in a ball containing $B_1$ and $B_2$. We can thus interpret this operator as a topological or non-local operator in two balls generating a ``generalized symmetry'' on the algebra of $R'=(B_1\cup B_2)'$. Similar non-local operators will appear in any region with non-trivial homotopy group $\pi_0(R)$, that is, a disconnected region.

In this scenario we could also consider regions $R'$ complementary to a disconnected region $R$, having non trivial homotopy group $\pi_{d-2}(R')$. If the model has a continuous Noether current $j_{\mu}^a$ associated with continuous global symmetry, we can integrate the conserved charge over the ball $B_1$   to produce a ``twist''
\be 
\tau_g =e^{i\int_{B_2'} \alpha_a j_0^a}\;,
\ee
where $\alpha^a$ are suitable smearing functions constant over $B_1$ and vanishing over $B_2$. If the group is non-Abelian we have to further average over group rotations to get a neutral element, which is then labeled by conjugacy classes $[g]$ of the group \cite{Casini:2020rgj,Casini:2019kex}.   
The twist can be defined as an operator that generates the symmetry transformation in $B_1$, but it does nothing in $B_2$. It therefore commutes with the additive algebra ${\cal A}(B_1)\vee {\cal A}(B_2)$ in $R$. But it will not commute with the intertwiners, which are charged in $B_1$.  Although the twists operators cannot be constructed by multiplying local operators in $R'=(B_1\cup B_2)'$, they can be constructed locally in a topological ball containing $B_1$. We can thus interpret this operator as a topological operator producing a ``generalized symmetry'' on the maximal algebra of $R=B_1\cup B_2$.  The twists $\tau_g$ for discrete symmetries are guaranteed to exist on general grounds, though it is more difficult to show their explicit form. Under certain assumptions, they can be constructed using modular theory, see \cite{Doplicher:1984zz}. But its existence is implied by one of the charged intertwiner operators because of von Neumann's double commutant theorem.   

For both generalized symmetries, one can find fusion algebras. The intertwiners fuse as the representations of the global symmetry group, see \cite{Casini:2020rgj} for the explicit construction, while the invariant twists fuse as the conjugacy classes do. There is of course the same number of conjugacy classes and representations. Indeed, for a given group, there cannot be two different theories with two different sets of representations. For each possible choice of the spectrum of representations, there is an associated group with a dual set of conjugacy classes. This is one corollary of the reconstruction theorem \cite{Doplicher:1990pn} described below.

One concludes that usual global symmetries correspond to violations of duality for regions with non-trivial $\pi_0 (R)$ and $\pi_{d-2}(R)$, and are naturally accounted in this framework.\footnote{Conventionally, it is said that twists generate the global symmetry operation while charged operators are just the non-invariant operators. In the present perspective, more duality symmetric, we could view as well the charged operators as effecting a generalized symmetry transformation under which the twists are charged themselves.} As a result, theories with global symmetries are not complete in the sense of this paper. 

One of the few known general results about the possible structure of non-local operator algebras is the reconstruction theorem \cite{Doplicher:1990pn}, see also \cite{Longo:1994xe}. It is based on the DHR approach to superselection sectors in QFT \cite{Doplicher:1969tk,Doplicher:1969kp,Doplicher:1971wk,Doplicher:1973at}. This theorem deals precisely with sectors associated with multiple balls. It is framed in a somewhat different language from the one of the present paper, more focused on superselection sectors in the space of states than on duality and additivity, and the authors make some additional assumptions. Under quite general conditions they can show that the above example is general enough in the sense that there is always a global symmetry group and a structure of bosonic or fermionic local charged operators $\psi^{i,r}$ behind sectors attached to disconnected balls. This result does not extend to $d=2$ where more complicated fusion rules are possible \cite{Doplicher:1971wk,cmp/1104179464,cmp/1104200513}. Examples of sectors with fusion rules not coming from a group include topological models in $d=3$, better described by the BF approach to superselection sectors \cite{Buchholz:1981fj}, see \cite{frohlich2006quantum} for a more throughout account.  However, these non-conventional sectors in topological models are attached to unbounded cone-like regions and end up being equivalent to conventional ones in $d=2$ models at infinity. In principle, they do not produce exotic sectors for localized regions in the $d=3$ theory in the topological limit.
 
Let us consider now gauge theories.  For any gauge theory with gauge group $G$, continuous or discrete, we have the well-known Wilson loops $W_r$, that is a set of gauge-invariant operators labeled by the representations of the gauge group. These operators are typically defined by the multiplication of non-gauge invariant operators over a loop. They are thus obvious candidates for non-local operators in regions with non-trivial homotopy group $\pi_1 (R)$, such as ring-like regions.

One of the main results in \cite{Casini:2020rgj} is that for $d\ge 4$ and any $r$, the Wilson loop in representation $rr*$, where $r*$ is the conjugate representation, can be locally generated inside the ring.\footnote{ We remark this is valid for any gauge group and $d\ge 4$, whether continuous or discrete, and its validity does not depend on how the QFT is defined in the UV, just on the existence of effective local physics.}
 Intuitively, for a Lie group, this is seen because of the existence of Adjoint Wilson lines. These are ended by the curvature tensor that can be used to break the Adjoint Wilson loop into pieces. But this result is more general as it can be applied to finite gauge groups as well. This result implies the operator $W_{rr*}$ is not topological and it cannot fail to commute with any operator linked with it. It does not give rise to a generalized symmetry. The same comments apply to all representations appearing in the fusion of $rr*$ with itself an arbitrary number of times. The reason is that Wilson loops can be chosen to satisfy the fusion rules of the gauge group
\be 
W_r W_{r'}=\sum\limits_{r''}\,N_{rr'}^{r''}\,W_{r''}\,.
\ee
If $W_{r}W_{r^*}$ can be locally generated inside the ring, then it follows that all Wilson loops appearing in the fusion of two such Wilson loops can be locally generated as well. 
More precisely, as described in \cite{Casini:2020rgj}, let us call the set of sectors appearing in arbitrary products of $rr*$ by ${\cal R}_{\textrm{add}}$, since they are all additive in the ring. This set of representations also coincide with the one generated by the adjoint representation of the group. By construction, ${\cal R}_{\textrm{add}}$ defines a subcategory of the category ${\cal R}$ of all representations. The true classes associated with the violation of duality arise when we quotient  ${\cal R}$ by  ${\cal R}_{\textrm{add}}$, which forms an Abelian group. In the literature of tensor categories, see \cite{etingof2016tensor}, this is called the universal grading of ${\cal R}$, and the associated group the universal grading group.  For our case, it is the group generated by the representations of the center of the gauge group and turns out to be isomorphic to the center of the group itself (Pontryagin duality). Therefore, the ``truly non-local'' Wilson loops run over the representations of the center.\footnote{A further comment about Wilson loops is the following. In the lattice or the continuum, they are usually defined as line operators. These line operators do not represent physical operators in the continuum and should be smeared. There are known difficulties to construct explicit smeared Wilson loops for non-abelian gauge theories. We note that there is no reason to expect that smeared physical Wilson loops exist for different representations $r$ of the gauge group because this label cannot be measured in any consistent way. On the contrary, smeared Wilson loops must exist for the different non-local classes determined by representations of the center. A construction using modular theory is described in \cite{Casini:2020rgj}. A construction using an extension to the non-gauge invariant Hilbert space is described in \cite{Pedro}. In such reference, Wilson loops are labeled by weights (points in the weight lattice). These weights have a definite ``N-ality'' which informs about how the center elements are represented in the representations with such weight.} These non-local operators are again topological and generate a generalized symmetry for the maximal algebra of the complementary region $R'$, which has non-trivial $\pi_{d-3}(R)$. We then end up in the framework of \cite{Gaiotto:2014kfa}, where the idea of generalized symmetries was first introduced.

As a result of this, one concludes that gauge groups with no center, whether discrete or continuous, \emph{do not} display generalized symmetries in dimensions greater than three. The reason is that in these cases, the fusion $rr*$ with itself produces all possible representations. This solves some of the issues raised recently in \cite{Harlow:2018tng,Rudelius:2020orz,Heidenreich:2021tna}, regarding special discrete groups with no center. Given such potential problems, it was argued in \cite{Rudelius:2020orz,Heidenreich:2021tna,McNamara:2021cuo} that further charged operators, apart from the ones needed to break the center symmetry, were needed to have a complete theory. The present arguments show this is not the case for such theories, since any consistent local effective field theory describing the gauge theory will have those ``adjoint'' charged operators \emph{built in} through the $rr*$ Wilson loops.

On the other hand, by integrating the gauge constraint for a central element $z$ of the gauge group, one can easily construct gauge invariant operators that violate duality in regions with non-trivial $\pi_{d-3}(R)$. These ``electric operators'', by opposition to the magnetic Wilson loops,  are the 't Hooft loops $T_z$ \cite{tHooft:1977nqb}. They form a group isomorphic to the center of the gauge group. Again, these operators are topological (since they can be transported and modified by using locally generated operators in the appropriate region). They thus generate a generalized symmetry for the additive algebra of the complementary region. Linked non-local Wilson loops and 't Hooft loops do not commute with each other. 

The special structure of dual Abelian groups for the fusion of dual non-local operators in gauge theories can be recovered abstractly without thinking about gauge theories. Assuming $d\ge 4$ and the absence of non-local operators for disjoint regions, in \cite{Casini:2020rgj} it was shown that non-local operators for a ring-like region and its complement always form Pontryagin dual Abelian groups, and their commutation relations are fixed as
\be
W_{z^{*}} T_{z}=\chi_{z^{*}}(z)\,T_z W_{z^{*}}\,,
\ee  
where $z\in Z$ is an Abelian group, $z^*$ is a representation of $Z$, and $\chi_{z^{*}}(z)$ is the character.

In $d=3$ ring-like regions are equivalent to the complement of two ball regions, and the topological difference between gauge and global sectors disappear. Fusion rules of non-Abelian groups are possible for non-trivial $\pi_1(R)$ in $d=3$.

Similar examples using gauge theories for $k$-form fields give rise to violations of duality in regions with non-trivial $\pi_{k}(R)$ and $\pi_{d-2-k}(R)$.  All the generalized global symmetries defined in \cite{Gaiotto:2014kfa} are included in this framework.  We expect sectors with non-group fusion rules to appear when combining generalized symmetries corresponding to different topologies in higher dimensions.

\section{Eliminating magnetic symmetries removes electric ones}
\label{em}

Given the existence of a generalized symmetry, we have derived the existence of a dual generalized symmetry. We thus expect the converse to be true as well, namely, when one of the symmetries disappears the other must go as well. We elaborate now on how this happens in the specific example of Wilson and 't Hooft loops and draw general lessons. 

By introducing some appropriate electric charges, the Wilson loops can become local operators. This is due to the now available Wilson lines ended by such charges, that can be used to decompose them into a product of operators along the path. 
 This automatically eliminates the electric symmetries generated by the dual 't Hooft loops,  without introducing magnetic charges to make those dual 't Hooft loops additive. The reason is that the electric charges that break Wilson loops also destroy the topological nature of the dual 't Hooft operators. As any other operator in the gauge theory, these operators are constructed additively in a ball, but can only be associated with a ring whenever they commute with everything local outside the ring. Once we introduce electric charges (which by definition are charged under the electric generalized symmetry), these new local operators do not commute with the 't Hooft operator and invalidate the association of the 't Hooft operator to a ring. In other words, it ceases to be topological, though it still exists in the algebra as a surface operator for a contractible surface. This observation has been recently elaborated independently in \cite{Casini:2020rgj,Heidenreich:2021tna}.
 
We arrive at a general lesson for the explicit breaking of a generalized symmetry. It can be summarized as:
\bigskip 
 
\noindent  {\sl If a non-local operator is made additive, the non-commuting dual non-local operators must cease to be topological.}  

\bigskip

\noindent However, since the theory is changed during the process of adding charges, this statement has sense only if the changes are perturbative or controllable, such as to allow the identification of the structures of the previous theory.    

Let us describe the general case for gauge theories. Starting with pure gauge theories in $d>4$ 't Hooft loops and Wilson loops have different dimensionality and live in complementary regions with different topologies. The dual symmetry groups are the center $Z$ of the gauge group and its dual $Z^*$ (group of characters of $Z$). Let us now explain how do these features change when one includes matter. Matter fields will break non-local operators if they are charged under the center $Z$ of the group (electrically charged fields that break Wilson lines) or the dual $Z^*$ of the center (magnetically charges fields that break 't Hooft loops).  Let us call $e\subseteq Z^*$ and $m\subseteq Z$ to the electric and magnetic charges with respect to $Z$ and $Z^*$ respectively. These can fuse, and then ``$e$'' and ``$ m$'' are subgroups. 

However, we are not allowed to introduce any charges with impunity. This restriction is a generalized Dirac quantization condition for explicit breaking of a generalized symmetry, and can be stated as: 

\bigskip

\noindent {\sl Two dual and non-commuting non-local operators cannot be simultaneously made additive}. 

\bigskip

\noindent Again this is a perturbative statement. The reason for this condition is that additive operators in complementary regions must commute by causality, and that was not the case for the original topological operators. 

In the present case of gauge theories, this gives $[m,e]=0$.   
 Now, all 't Hooft operators in $Z$ not commuting with $e$ can no longer be considered topological operators living in a ring. They are now the only operators that exist in balls. Therefore, the remaining t 'Hooft loops in the ring are given by $(Z^*/e)^*\subseteq Z$, which is a subgroup of $Z$.\footnote{In group theory terms, if $e$ is a subgroup of $Z^*$, then we define $(Z^*/e)^*$ to be the subgroup of $Z$ formed by all the elements $z$ for which $\chi_{e}(z)=1$, so that the associated operators in this subgroup commute with the operators in $e$, and the Dirac-quantization condition is satisfied. Analogously, if $m$ is a subgroup of $Z$, then we define $(Z/m)^*$ to be the subgroup of $Z^*$ formed by all the elements $z^*$ for which $\chi_{z^*}(m)=1$ so that the associated operators in this subgroup commute with the operators in $m$.} The subgroup $m$ has to be included in this subgroup due to the Dirac quantization condition. This implies that the loops in $m\subseteq (Z^*/e)^*$ are now locally generated, broken by magnetic charges. Hence, the remaining generalized electric symmetry, generated by the remaining non-local 't Hooft loops in the ring is given by $(Z^*/e)^*/m$. Dually, and for equivalent reasons, the non-local Wilson loops will be $(Z/m)^*/e$. Notice that
\be 
(Z^*/e)^*/m\simeq (Z/m)^*/e\;.
\ee
Therefore, at the end of the day, the remaining generalized symmetries have still the same size and correspond to dual Abelian groups. In other words, we see that whenever we break a certain amount of generalized electric symmetry, we break an equal amount of generalized magnetic symmetry.

In $d=4$ both symmetries live in the same topology of rings. Therefore, without matter fields, the generalized symmetry is $Z^*\times Z$. The first $Z^*$ is generated by the non-breakable Wilson loops and the second isomorphic group $Z$ is generated by the non-breakable 't Hooft loops. To make additive some operators one can introduce a group $D\subseteq Z^*\times Z$ of dyons, having electric and magnetic charges. Analogous reasoning gives $((Z^*\times Z)/D^*)^*/D$ as the remaining group of symmetries.

We conclude that, in parallel to the statement that generalized symmetries always come in pairs, together with its precise quantification via index theory and complementarity diagrams, we have that breaking a certain amount of some symmetry implies breaking the same amount of the dual symmetry.

\section{The certainty relation and the fate of symmetries with the scale}

Another form in which a symmetry may disappear is in an effective manner, considering the physics of the theory at different scales. One may wonder if our statement that generalized symmetries come in dual pairs survive this consideration. More concretely, one wonders if a symmetry may be visible in the IR, while the other dual one only at the UV, making the assertion that symmetries come in dual pairs of little physical significance. We now argue that this is not the case. 
 
There are two subtleties to consider. First,  the topologies of the regions where the non-local operators live generally have geometries determined by several length parameters. Therefore, to consider the behavior with the scale we have to make an overall scaling, keeping fixed the shape of the region. The second is that non-local operators inside these regions can themselves have different support. A `thin'' operator or, as a limit, a line operator, has no definite scale, and always has vanishing expectation value. To fix a scale we have to take a smeared operator, with smooth smear functions, filling the region $R$. 

Let us take a simple example with a global symmetry and where the charged operators become massive in the IR. The intertwiner will have zero expectation value for large regions and distances, and we can rightly say that it will effectively disappear in the IR. On the other hand, in the IR, the twist can be smeared on a scale larger than the typical mass of the charged particles, in such a way that massive charge-anticharge vacuum fluctuations fail to produce an effective charge on each side of the localized twist. In that case, it is not difficult to see that $\langle \tau\rangle \rightarrow 1$, as the twist acts quite similarly to the global group operator. This smooth twist essentially leaves all the low energy states unchanged, which is not surprising, given that charged operators are very suppressed. The opposite happens when there is spontaneous symmetry breaking (SSB). The charge-anticharge operator, with adequate smearing, will set to constant expectation value in the IR. The large twist will have almost zero expectation value because it connects a vacuum with an orthogonal one. In this light, these two regimes, massive charges for unbroken symmetry, and SSB express dual phenomena in regards to the generalized symmetries. The same also happens for the SSB and confinement in gauge theories \cite{tHooft:1977nqb}. In more generality, this is how a pair of dual symmetries  disappear: 

\bigskip

\noindent{\sl Dual symmetries disappear together along the RG (or do not disappear). When non-local operators of one symmetry fall out of the spectrum of accessible operators the dual operators can be assimilated to numbers.}

\bigskip  

Instead of using non-local operators, which are subject to large ambiguities in their choice, there is a more suitable tool to discuss these issues. 
This involves the relative entropy between two states. Relative entropy measures the distinguishability of two states by physical operations in a precise operational way \cite{ohya2004quantum,petz2007quantum}. Here the relevant relative entropy, which we have called {\sl entropic order parameter} of the corresponding symmetry,  is  
\be
S_{{\cal A}_{\rm max}(R)}(\omega|\omega\circ E)\,.
\ee
This compares, in the algebra ${\cal A}_{\rm max}(R)$, two states, namely the vacuum $\omega$, and the vacuum composed with the conditional expectation $\omega\circ E$. This last state kills the non-local operators. Therefore, this relative entropy is a measure of the statistics of non-local operators. But, importantly, it does not rely on any specific choice of the non-local operator. In a loose sense, it automatically chooses the ``best ones'', namely the ones with the largest expectation values, and it does not forget the additive operators either. This relative entropy is bounded by the logarithm of the index 
\be
0\le S_{{\cal A}_{\rm max}(R)}(\omega|\omega\circ E)\le \log \lambda\,,\label{lala}
\ee
where $\lambda=|G|$, the size of the group, for the examples we have discussed. It approaches the lowest bound when non-local operators have zero expectation value and the upper bound when they saturate to their maximal expectation value. 

In comparing the statistics of dual non local operators we have the {\sl entropic certainty relation} \cite{Casini:2019kex,Magan:2020ake,Casini:2020rgj}  (see also \cite{hollands2020variational,xu2020relative}) for a global pure state:
\be
S_{{\cal A}_{\rm max}(R)}(\omega|\omega\circ E)+S_{{\cal A}_{\rm max}(R')}(\omega|\omega\circ E')=\log \lambda\,.
\ee
This relation tells the two bounds on (\ref{lala}) are achieved in a complementary way by the two dual order parameters. Both bounds can \emph{actually} be achieved in the theory. The reason is that since the lower (vanishing) bound of the entropic order (disorder) parameter is always achieved by letting the region going to zero internal size, then the upper bound of the entropic disorder (order) parameter is achieved in this same limit. The certainty relation contains information on the uncertainty relations for the non-commuting dual non-local operators. But it is more powerfully expressed as an equation rather than an inequality.  

Returning to the analysis of the fate of dual symmetries along the renormalization group, it is clear that as the regions are scaled keeping the shape constant we can get saturation of one relative entropy at the expense of the vanishing of the other. In one case, smeared non-local operators will saturate to maximal expectation values while the dual ones will have zero expectation values. These last ones disappear at low energies and the former will act as numbers. If this saturation is fast enough in relation to the scaling parameter, correlations between non-local operators become trivial, which is a manifestation of a ``gap'' in the theory.    Insightful examples of saturated dual symmetries are topological limits. These limits are controlled by topological field theories, in which 't Hooft loops decouple and the expectation value of localized Wilson loops saturate to one.

\section{Final comments}

We have shown that the analysis of the basic properties determining the way algebras are assigned to regions in QFT gives a simple and unifying understanding of completeness, the origin of generalized symmetries, the existence of dual symmetries, and the idea of completeness as the absence of generalized symmetries. We hope this understanding could be the basis of a classification of possible generalized symmetries for different space-time dimensions and regions. In what follows we discuss topics that have attracted attention or produced confusion in the literature concerning the idea of completeness, and briefly look at how they fit in the framework of the present paper.    

\subsubsection*{Modular invariance and completeness}

Modular invariance of $2d$ CFT's is sometimes invoked by simplicity or aesthetic reasons and sometimes assumed as a new axiom, though without compelling physical reasons. In string theory (worldsheet formulation) it appears as a fundamental requirement.  The $T$ transformation of the modular symmetry is related to mutual locality, or fermion locality, between the fields and this is necessary for a local, real-time, QFT. The $S$ transformation interchanges the axes of the torus and is not mandatory. It does, however, imply an important relationship for the entropies through the replica trick \cite{Cardy:2017qhl}. It follows from modular invariance that the Renyi entropies of two intervals are equal to the Renyi entropies of the complementary two intervals (in a circle). This is by no means automatic, because Renyi entropies in a global pure state are equal for dual algebras $S_n({\cal A})=S_n({\cal A}')$ rather than complementary regions. The equality of entropies for complementary sets of two intervals is a necessary condition for duality for the two-interval algebras. In the path integral formulation, where these calculations are performed, it is clear that the interval algebras are generated locally. Consequently, we have additivity and duality, at least for two interval sets, for the algebras defined by the path integral.

Therefore modular invariance is a form of completeness for a $2d$ CFT in a sense similar, but weaker, to the one we have discussed above because the additive algebras are not required to be observable algebras. A mathematical analysis of CFT models and modular invariance, and its relation with the trivial index for two intervals, can be found in \cite{xu2020relative}. From the physical point of view, modular invariance implies that in a torus where one direction represents time and the other space we could freely interchange the role of the two directions. Non-local operators for two intervals can be pictured as flux lines at $t=0$. If put in the time direction this same line represents a charged particle world line, and these charges are the ones that make the non-local operators additive.

From this perspective, the notion of completeness we have described extends and generalizes the notion of modular invariance to higher dimensional CFT's.

\subsubsection*{Haag-Dirac nets vs completeness}
\label{dvh}

A {\sl net} is an assignation ${\cal A}(R)$ of algebras to regions satisfying isotonia \eqref{isotonia} and causality \eqref{causality}. 
The choice of additive algebras ${\cal A}_{\rm add}(R)$ for any $R$ is always a net, but for non-complete theories, the choice of maximal algebras ${\cal A}_{\rm max}(R)$ for any $R$ does not form a net since complementary non-local operators do not commute. However, for non-complete theories, we can always construct different nets, which are based on the same underlying theory. We can form a net by adding non-local operators to the additive net, being careful not to add non-commuting operators to complementary regions.  A maximal net in this sense is necessarily one that satisfies duality
\be
{\cal A}(R)=({\cal A}(R'))'\;,
\ee
for every $R$, where ${\cal A}_{\rm add}(R)\subseteq {\cal A}(R)\subseteq {\cal A}_{\rm max}(R)$. But, of course, such a net will not satisfy additivity. In general, there are many different incompatible ways to choose a maximal net satisfying duality. We have called these nets Haag-Dirac nets because for gauge theories the solutions are directly related to solutions of the Dirac quantization condition, see \cite{Casini:2020rgj}. This is a non-perturbative way to arrive at the Dirac quantization condition, and any of these nets (satisfying additionally all spatial symmetries) could in principle be used as a starting point for an explicit breaking of the symmetries, by making the non-local operators additive,  as described in section \ref{em}.   

However, we emphasize all Haag-Dirac nets of a given non-complete theory are physically equivalent.  This is a point that gives place to confusion from time to time.  The theory is not changed by the way we choose to associate non-local operators to regions. These non-local operators already necessarily form part of the additive algebras of balls.  In particular,  there is no completeness for these Haag-Dirac nets, because completeness happens when the duality property is satisfied by the additive net, which then becomes the only possible net.   

\subsubsection*{On a dense set of charges charges}
\label{irr}

There is a model that might be exhibited as a counterexample of our assertion that all generalized symmetries come in dual pairs, and that has appeared in the literature to argue for charge quantization in quantum gravity. The idea is to think in a $U(1)$ gauge symmetry for example and choose a dense set of charges, say rational charges $q\in {\cal Q}$. The fusion does not take the charges out of this set. Then, all Wilson loops of rational charge are additive, but in principle, the non-rational charges are associated to a group of generalized symmetries ${\cal R}/{\cal Q}$, which is still dense.  Therefore, there cannot be any 't Hooft loop. Such loop of magnetic charge $g$ would have a commutation relation 
\be
W_q T_g= e^{i q g}\, T_g W_q\,,
\ee     
for all the additive loops with $q\in Q$. It is not possible that the phases cancel for all rational $q$, and in consequence, the 't Hooft loop is not a non-local operator but a surface operator. 

The puzzle is that, as we described, dual symmetries are implied robustly by von Neumann's double commutant theorem. The solution to this puzzle is that this theorem holds only for von Neumann algebras, which are closed under weak limits. Then it clearly must be the case that the dense set of charges generate all charges under weak limits. This shows the physical impossibility of this model since any rational set of charges could not be physically distinguished from the real line.

This model was considered in \cite{Heidenreich:2021tna}. Given the previous observation about dense sets of charges, it was claimed that the equivalence between completeness and the absence of generalized symmetries fails for this and other similar models. As the previous argument shows this is not the case. We conclude that the equivalence between completeness and the absence of generalized symmetries is still valid for dense sets of charges, as ensured by von Neumann double commutant theorem.

\subsubsection*{Completeness in holographic theories of quantum gravity}
\label{holo}
In the AdS/CFT or holographic duality \cite{Maldacena:1997re}, one of the most famous and old entries of the holographic dictionary is the relation between gauge symmetries in the bulk and global symmetries at the boundary \cite{Witten:1998qj}. At the technical level, the gauge field $A_\mu$ is dual to the global CFT current, which means that the algebra of such current is generated by the boundary limit of the gauge field. This is called the extrapolated dictionary, see \cite{Banks:1998dd}. This current, as we have described before, can be integrated over finite balls to produce non-local operators. These are the twists and violate duality in regions with non-trivial $\pi_{d-2}$.

Although the previous picture might hold at the level of generalized free fields, if the dual theory is a  real CFT (with a stress tensor generating time evolution), the previous picture cannot be the whole story. As we learned above, whenever we have some non-local operators violating duality in a region $R$, von Neumann's double commutant theorem forces the existence of dual non-local operators violating duality in the complementary region $R'$. In the previous case, this means there must exist intertwiners (charge-anticharge operators) in all representations of the global symmetry group violating duality in regions with non-trivial $\pi_{0}$. These are generated by local fields of certain scaling dimensions. These fields are dual to massive bulk charged fields. Since there are intertwiners in all representations of the global symmetry group, we conclude that all bulk Wilson loops can now be broken into pieces. This implies that there is no duality violation in the bulk and the bulk effective theory cannot have $1$-form symmetry incompleteness. One nice feature of this argument is that it extends to higher form symmetries in the bulk, dual to higher form symmetries in the boundary. It also potentially extends to generalized symmetries not originated from a group.

Arguments against the existence of generalized symmetries using AdS/CFT have been given in recent literature, see \cite{Harlow:2018tng,Harlow:2015lma,Harlow:2021trr}. We remark that the present argument uses only one boundary, and it is constructive in the sense of not being based on any assumed contradiction or potential problem. The holographic nature of quantum gravity, together with basic principles of QFT, such as von Neumann's double commutant theorem, suggests that ``electrons'' (charged fields) and ``photons'' (gauge fields) in the bulk are, in a holographic sense, dual to each other.

\section*{Acknowledgements} 
The authors are indebted to many discussions with M. Huerta and D. Pontello 
  which lead to the works here reviewed. 
The work of H. C.  is partially supported by CONICET, CNEA, and Universidad Nacional de Cuyo, Argentina, and an It From Qubit grant by the Simons Foundation. The work of J.M is supported by a DOE QuantISED grantDE-SC0020360 and the Simons Foundation It From Qubit collaboration (385592).

\bibliography{EE}{}
\bibliographystyle{utphys}



\end{document}